# Eight Transaction Papers by Jim Gray

Philip A. Bernstein
Microsoft Research
October 6, 2023

**ABSTRACT**

This article is a summary of eight of Jim Gray's transaction papers. It was written at the invitation of Pat Helland to be a chapter of a forthcoming book in the ACM Turing Award winners' series, *Curiosity, Clarity, and Caring: How Jim Gray's Passion for Learning, Teaching, and People Changed Computing.*

**INTRODUCTION**

Jim Gray's first big research success was in defining and popularizing the transaction abstraction and techniques for implementing it. It was also a major component of his Turing Award citation: "for seminal contributions to database and transaction processing research and technical leadership in system implementation". He started writing about the topic in the mid-1970's, and periodically revisited it for the next 30 years. His 1993 book, *Transaction Processing – Concepts and Techniques*, coauthored with Andreas Reuter, is a classic [9]. It is still used by today's implementers to guide them on how best to implement transactions in a database system.

This chapter selects some of Jim's most impactful papers. They describe his many research contributions and the evolution of his thinking on the topic.

With his many collaborators, Jim created transactions as one of the foundational abstractions of software. He also promoted it as a research field. I spent the early part of my career following in his footsteps. Like Jim, I have revisited the topic many times since then. It is an honor to have been asked to summarize Jim's work on transactions. I hope I have done it justice.

_______________________________

Paper #1: Kapali P. Eswaran, Jim Gray, Raymond A. Lorie, Irving L. Traiger: ***The Notions of Consistency and Predicate Locks in a Database System.***
*Commun. ACM 19(11): 624-633 (1976)*

The transaction abstraction is well known to everyone who studies or uses database systems. What is not so well known is that the main concepts of the abstraction were defined in a single research paper by Jim and his colleagues. This landmark paper introduced the transaction abstraction and the most popular way to implement transactions, namely two-phase locking. It also set the stage for decades of transaction research.

The paper made four huge contributions:

1. It defined a transaction to be a sequence of operations over a shared state comprised of "entities" (e.g., records of a file or tuples of a relation). It assumes that a transaction preserves the internal consistency of that state.

2. When executing transactions concurrently, the correctness goal is that the actual execution should be equivalent to a serial (i.e., non-interleaved) execution of the same transactions. Since the serial execution obviously preserves consistency of the state, the actual (concurrent) execution does too.

3. It invented the two-phase locking protocol, which defines three rules for how transactions should go about locking shared state.
    i. A transaction acquires a lock on each entity before accessing it.
    ii. A transaction holds its lock on an entity until after it is done accessing the entity.
    iii. A transaction acquires all the locks it needs before it releases any of them.

It showed that an execution of transactions that uses two-phase locking is *consistent*, in the sense that the execution will have the same effect as a serial execution of those transactions. Today, we call this consistency property *serializability*.

4. It showed that operations that retrieve a set of records based on their field values cause a problem for two-phase locking, called the *phantom problem*. They proposed solving it using *predicate locks*, each of which identifies a set of data items to be locked based on their values rather than by their identities.

Before we delve into these contributions, consider the state of the art when this paper was written. It was well known that concurrent read and write operations by different programs on the same data can lead to inconsistencies. One example is a race condition, such as when two concurrent programs add 1 to a shared variable x, but they both read x before either of them writes to x, so one of the updates gets lost. Another example is inconsistent reads, where one transaction moves $100 from $Account_1$ to $Account_2$, and another transaction reads $Account_1$ before the transfer and $Account_2$ after the transfer. It was understood that setting locks on shared data was a way of avoiding such undesirable outcomes. But it was up to the application program to set and release these locks, with little guidance about exactly which locks should be set and when.

This paper provided that guidance. It said that:

1. Read and write operations should be grouped into consistency-preserving units, called transactions. (Contribution 1)
2. The correctness goal is to ensure every allowable execution is equivalent to a serial execution of transactions. (Contribution 2)
3. A way to guarantee the correctness goal is to follow the two-phase locking protocol. (Contribution 3)
4. There is an important problem that arises when retrieving a set of records based on their field values. (Contribution 4)

The problem in contribution 4 can be understood from the following example, which they presented. Consider a database consisting of an Accounts table with three fields [Account#, Location, Balance], and an Assets table with two fields, [Location, Total]. The database is consistent if the sum of Balance values of all accounts for a given location in the Accounts table is equal to the Total value for that location in the Assets table. Consider the following sequence:

i. An audit transaction $T_1$ checks the consistency of data with Location = 'Napa'. To do this, $T_1$ locks all the rows in the Account where Location is 'Napa'.
ii. Another transaction $T_2$ inserts a new row in Accounts with Location = 'Napa' and adds that account's Balance to Total in the row in Assets where Location = 'Napa'. $T_2$ terminates and releases its lock on the two rows that it accessed.
iii. $T_1$ continues executing by locking the row in Assets where Location = 'Napa'. However, when it compares the value of Total in that row to the sum of account balances that it read in step (i), it finds they are not equal.

This example is puzzling because both transactions are two-phase locked, yet the effect of the execution is different from running $T_1$ and $T_2$ serially in either order. This arises because $T_2$ is a phantom row, i.e., it comes and goes like a ghost. This seems to contradict the theorem that proves two-phase locking ensures serializability. The puzzle is solved by recognizing that there is a hidden operation in step (i), namely, the method by which $T_1$ determined which rows of Account have Location = 'Napa'. That operation might have accessed an index that maps each value of Location to all rows with that Location. Or it might have scanned the table, checking each row for Location = 'Napa', until it hit the end-of-table marker. Whatever technique is used, there is a data item, such as the index or end-of-table marker, that $T_1$ must have accessed to identify all rows with Location = "Napa", and $T_1$ must have locked that data item. To insert a row in Account with Location = 'Napa', $T_2$ must have updated that data item too. However, since it would have been prevented from doing so by $T_1$'s lock, the above execution sequence could not have happened.

To avoid this phantom problem, the paper suggests setting locks on predicates, such as "Location = 'Napa'", or Boolean combinations of these predicates, such as (((Location = 'Napa') or (Location = 'Santa Rosa')) and (Balance < 200)). It calls them *predicate locks*. To set a predicate lock, a transaction must check that no other transaction owns a predicate lock on the same table that is mutually satisfiable with the requested lock. Unfortunately, such checks are in general computationally expensive. In the next paper, they propose a more efficient alternative.

Paper #2: Jim Gray, Raymond A. Lorie, Gianfranco R. Putzolu, Irving L. Traiger: **Granularity of Locks and Degrees of Consistency in a Shared Data Base.**
IFIP Working Conference on Modelling in Data Base Management Systems 1976: 365-394

If every transaction locks only the records it accesses, transactions that operate on different records can run concurrently. However, some transactions access most or all records in a file. Jim and his colleagues recognized that it would be prohibitively expensive for such transactions to lock every record one-by-one. Fortunately, it is unnecessary to do so. Since a transaction that locks all records of a file prevents other transactions from operating on any record of the file, it might as well set only one lock on the file instead of many locks on the individual records. That addresses the problem of locking expense. However, if a transaction $T_1$ locks the file and another transaction $T_2$ locks individual records, how does the system detect that these locks conflict? The answer proposed in this paper is *multigranularity locking*, which is now a standard solution used in relational database systems and many others. The idea is to require that before a transaction locks a record, it sets a weak lock on the file, called an intention lock, that warns other transactions against locking the entire file.

This paper, like all later papers and systems, exploits the fact that read operations do not interfere with each other and therefore should be allowed to run concurrently. It therefore distinguishes *Share* (S) locks for read operations and *Exclusive* (X) locks for write operations. Many transactions can simultaneously hold an S lock on an entity *e*, thereby allowing many readers of *e*. However, a transaction T can hold an X lock on *e* only if no other transaction has a lock on *e,* thereby ensuring T has exclusive access to *e*.

Multigranularity locking organizes lockable entities into a hierarchy. For example, a *database* contains *areas* (e.g., disk volumes), which contain *files*, which contain *records*. A lock on an entity, such as a file, implicitly locks all the entity's descendants, e.g., the records in that file.

It is important to prevent a transaction from holding an S or X lock on an entity *e* (e.g., a file) while another transaction holds an X lock on a descendant *d* of *e* (e.g., a record). To enable this, multigranularity locking introduces a new access mode, called *intention mode*.

A transaction T's intention-mode lock on an entity *e* indicates that T is acquiring finer granularity locks on descendants of *e.* It tells other transactions not to set certain kinds of locks on *e*, because those locks will conflict with T's finer granularity locks beneath *e* in the hierarchy. There are two types of intention locks, *Intention Share* (*IS*) and *Intention Exclusive* (*IX*). An IS lock on *e* says its owner has S locks on finer granularity entities beneath *e* in the hierarchy. For example, an IS lock on a file entity says that its owner has (or will have) S locks on records within the file. Therefore, an IS lock on *e* is incompatible with an X lock on *e* and therefore prevents another transaction from setting an X lock on *e*. For example, a transaction's IS lock on a file prevents another transaction from setting an X lock on the file, because that X lock would conflict with the S locks on records held by the owner of the IS file lock. Similarly, an IX lock on *e* is incompatible with an S or X lock on *e*. It prevents another transaction from setting an S lock or X lock on *e*.

The paper develops these lock modes in more detail. It introduces the SIX lock mode, which gives the owner shared access to the entity and the right to set X locks on the entity's descendants. This is useful when a transaction reads all the records in a file and updates some of them. The paper also explains what to do when the lock hierarchy is a directed acyclic graph, not a tree. This arises when a record can be accessed either by scanning a file or by lookups in an index (e.g., an index on the department field of the Employee file). The problem is that a transaction that accesses a range of index values (e.g., employees in the toy department) does not want to set a lock on the entire file. Moreover, a transaction might update a record, which moves the record from one index range to another (e.g., from the toy department to the clothing department), potentially creating a phantom problem. There are many interesting issues here, which you will learn by reading the paper.

The idea of multigranularity locking has been hugely influential. Most SQL database systems use it, and there have been numerous technical papers to fine-tune the technique. Jim and his coauthors published their description of multigranularity locking as a separate paper at the first International Conference on Very Large Data Bases (VLDB) in 1975 (a conference that has been held every year since then) [8]. Six papers from that conference were selected as the best papers of the conference, published in the first issue of ACM

Transactions on Database Systems, March 1976. In retrospect, it is surprising that the multigranularity locking paper was not one of them, in that only one of the other selected has been as influential (Peter P-S Chen's "The entity-relationship model—toward a unified view of data"). It is an example of how difficult it can be for a conference program committee to predict which papers in their conference will end up being most important.

The second part of the paper describes another aspect of transaction behavior, namely, when a transaction's updates become permanent. A transaction terminates by issuing either an abort or commit operation. The *abort* operation discards all the transaction's updates. The *commit* operation causes the transaction's updates to be installed permanently, thereby giving up the transaction's right to discard the updates.

Data that is written by a transaction but not yet committed is said to be *dirty*. Problems arise when a transaction is allowed to read or overwrite dirty data. Solving these problems leads to a hierarchy of locking protocols, which this paper calls *degrees of consistency*. Today, we call them *isolation levels*.

At degree 0 a transaction T holds an X lock on an entity only while performing its update. Thus, T releases the lock before it commits. They call this a *short duration lock*. Degree 0 can produce all sorts of problems when transactions abort. Suppose a transaction $T_2$ read a dirty value of entity *e* that $T_1$ wrote. If $T_1$ aborts, then $T_2$ should abort too, since it read data that is no longer valid. Moreover, it is important that $T_2$ does not commit until after $T_1$ commits. Otherwise, if $T_1$ aborts, $T_2$ is in a catch-22 situation. Since it committed, its updates are supposed to be permanent. But since it read invalid data, it is supposed to abort.

Another nasty situation arises if $T_2$ overwrites a dirty value of entity *e* that $T_1$ wrote. In this case, if $T_1$ aborts, then both $T_1$'s update and $T_2$'s update should be discarded. If $T_1$ commits and $T_2$ aborts, then only $T_2$'s update should be discarded. If there is a long chain of uncommitted updates to *e*, then some complicated bookkeeping is required to figure out what to do when one of the transactions aborts.

At degree 1 a transaction T sets an X lock on an entity before updating it and holds the lock until after T commits. This is called a *long duration lock*. It ensures a transaction cannot overwrite dirty data, which avoids one of the problems with abort. But reading uncommitted data is still possible.

Degree 2 solves the latter problem by adding another requirement to Degree 1, namely, that a transaction must get a short duration S lock on an entity before reading it. This ensures the transaction reads only committed data.

Degree 2 still leaves open another problem, namely, that the transaction might read some but not all of another transaction's updates. For example, suppose $T_1$ updates two entities, $e_1$ and $e_2$. $T_2$ could read the value of entity $e_1$ before $T_1$ updated it and the value of $e_2$ after $T_1$ updated it. To prevent this, Degree 3 adds the requirement that a transaction gets a long duration S lock on data it reads. With Degree 3, a transaction is two-phase locked and therefore (modulo the phantom problem) is serializable.

The paper explains all this in detail with some correctness arguments. It closes with a summary of locking behavior in two database systems that were popular at the time (1976), IBM's IMS/VS and UNIVAC's DMS 1100.

Paper #3: Jim Gray: **A Transaction Model.**
ICALP 1980: 282-298. Originally published as IBM Research Report RJ2895 (36591) August 7, 1980.

After Jim's early publications about transactions in 1975-1976, computer science researchers recognized that the transaction abstraction was a complicated concept that could benefit from a mathematical development. By 1979, several technical articles were published along these lines, such as [2,12,20,23,27].

Although Jim's primary interest was in building practical systems, he had a strong background in theoretical computer science. His early papers with his Ph.D. advisor at University of California Berkeley, Michael Harrison, were on automata theory and formal languages. He used his mathematical skill to define mathematical models of transactions, such as the proof that two-phase locking produces serializable executions. He decided to consolidate these models and extend them in this paper, which appeared as an IBM Technical Report and later published at the 7[th] International Conference on Automata, Languages, and Programming.

One extension was a model of logging and recovery of a transaction system. The paper distinguishes three types of entities: real entities, whose values cannot be changed, such as printed output; stable entities, whose values can be changed and are durably stored (e.g., on disk) and hence should survive system restart; and volatile entities, whose values can be changed but are not durable and hence do not survive system restart. It then discusses two types of failure actions: transaction failure, where the transaction loses its state and has to be restarted; and system failure, which causes the loss of all volatile entities, such as the internal state of all executing transactions, and therefore requires the system to be restarted. The system can protect itself from such failures by logging operations that modify state. After restarting a transaction or the system, a recovery procedure can redo the logged updates to return the system to its state shortly before the failure. The paper defines the formal conditions that a checkpointed state and redo actions must satisfy for the recovery procedure to return the system to its correct state. Although some previous articles had explained practical procedures to perform this kind of recovery, this was one of the first (if not the first) to formally define correct behavior.

Another extension was an intuitive model of transaction performance, based on the rate that transactions executed conflicting operations on the same data item. The paper showed that the probability a given transaction experiences a deadlock is linearly proportional to the number of transactions executing concurrently (a.k.a. the degree of concurrency). Therefore, the probability that some transaction deadlocks (and hence the deadlock rate) is proportional to the square of the degree of concurrency. This explains why it is important to limit the number of transactions that execute concurrently.

Finally, the paper models the problem of ensuring the atomicity of transactions that execute in a distributed system. By atomic, we mean that either all the transaction's updates are installed or none of them are. The solution to this problem is the two-phase commit protocol, published several years earlier by Butler Lampson and Howard Sturgis in a Xerox PARC technical report and later in anthology [15]. The main idea is that a transaction must save its updated results in persistent storage at all nodes where it executed before it commits at any node. That way, even if it fails after having committed at some nodes but not at others, there is enough information in persistent storage to complete the commitment process. In this paper, Jim gave a formal characterization of the problem and how two-phase commit satisfies it.

Jim was convinced that transactions are a fundamental abstraction and should be embodied in programming languages. The paper closes with thoughts about how to do this. Several years later, another Turing Award winner, Barbara Liskov, proposed such a language, called Argus [16].

Paper #4: Jim Gray: **The Transaction Concept: Virtues and Limitations (Invited Paper).**
VLDB Conference 1981: 144-154

By 1981, the transaction abstraction had been around long enough that its virtues and limitations were becoming more apparent. Jim summarized them in this paper, which covers a lot of ground: the transaction abstraction, fault tolerance, transaction implementation techniques, and extensions to the transaction concept.

The paper starts by relating the transaction concept to contract law, where two parties make a binding agreement. This leads to defining the properties of a transaction as consistency, atomicity, and durability. As in earlier papers, consistency subsumes isolation (i.e., serializability).

The previous paper ("A Transaction Model") discussed properties of entities. This one focuses on actions on entities. Some actions are *unprotected*, meaning they need not be undone if the transaction aborts. Others are *real*, meaning they cannot be undone, at least not by the computer, such as dispensing money in an ATM. Finally, there are *protected* actions, which are undone if the transaction aborts and are durable if the transaction commits. In some implementations, protected actions that are committed must be redone if the system fails after having committed the actions' transaction and before storing the actions' results in the persistent database.

After a transaction is committed, the only way to change its effects is to run another transaction, called a *compensating transaction*. This paper seems to be the first to have introduced this important concept.

By the time Jim wrote this paper, he had moved from IBM to Tandem Computers, a startup computer company that specialized in fault tolerant computers for on-line transaction processing. (The company is gone, but its products are still sold by Hewlett-Packard Company.) This got him thinking more about the relationship of transactions to failure handling. The paper talks about mean time between failures and mean time to repair and shows their effect on availability of hardware and software. As a hardware example, it shows how mirrored disks can greatly increase disk availability, especially if disk failure is detected quickly (a.k.a. *fail fast*). Software can be made fault tolerant by having a backup process for each process, provided the backup can continue the work of the failed process exactly where the latter left off. However, this requires that the recovery activity synchronizes communicating processes to mutually consistent states, which can be quite difficult. The transaction concept helps avoid this synchronization challenge because the system only needs to ensure that after recovering from failure, the effects of committed transactions are durable and uncommitted transactions are aborted and their effects wiped out.

The paper then summarizes two techniques for implementing atomicity and durability: *time-domain addressing* (which we now call multiversioning) and logging. With multiversioning, each transaction creates a new version (i.e., copy) of each entity it updates. Multiversioning can be integrated with consistency maintenance (i.e., concurrency control) by assigning each transaction a timestamp and associating that timestamp with each version the transaction writes. For a given timestamp *t*, a consistent state at time *t* consists of the version of each entity with largest timestamp less than or equal to *t*. The paper compares the pro's and con's of this approach with those of logging (and locking-based concurrency control), where the system keeps just one version of each entity that transactions update in place. At the time of this paper, logging was the more popular technique. Today, the techniques are frequently combined to enable recent (but not the latest) versions to be read, so that consistent read-only transactions can run concurrently with update transactions.

The paper closes with a discussion of three extensions that at the time were considered beyond the state of the art: nested transactions, long-lived transactions, and transactions integrated with programming languages. In a *nested transaction*, each transaction can have sub-transactions that are isolated from each other and atomic but are committed together by the parent transaction. An example is reserving a travel itinerary comprised of three subtransactions for flight, hotel, and car rental reservations.

Some transactions are *long-lived*, requiring hours to execute, such as settling an insurance claim. Two-phase locking does not work well here, since data might be locked for weeks while the insurance company, service provider, and client negotiate. His proposed solution is to allow only "active" transactions to hold locks.

Given the subtleties of the transaction concept discussed in this paper, it makes sense to expose them in the programming language using elegant abstractions with clean semantics. For example, a procedure could be tagged with the keyword *transaction*, thereby requiring that it transparently starts a new transaction on invocation and commits when it finishes. The three extensions described in this last section of the paper were among the main agenda items for the transaction field for the next 25 years.

In his invited talk at the SIGMOD 2006 conference entitled "Transaction Research – History and Challenges", Jim revisited many of the points he made in this 1981 paper.

In early implementations, a transaction overwrote the value of each entity it updated. This made sense when storage was expensive. By 2006, storage was much cheaper, and many systems adopted multiversioning. He likened this to generally accepted accounting principles, which do not allow in-place updates or deletes. You can only add information, e.g., that an entity has been updated or deleted.

Early implementations of transactions used a log to record updates and thereby ensure durability. Thus, the database is a replica of the log. By 2006, there were many replication technologies in common use. A new challenge is how to simplify the many options for replication.

For long-lived transactions, the field has settled on the concept of a "workflow", which consists of multiple transactions, but not on a specific software abstraction that embodies the workflow concept. Jim highlighted three concepts that have been useful in workflow systems: compensations, simple workflow structures, and commit-abort dependencies.

- Compensations are explained earlier in this section.
- One simple workflow structure is "sagas", where the transactions execute in sequence, one after another [18]. If one of the saga's transactions aborts, then the system invokes compensations for earlier transactions in the sequence, which already committed.
- Commit-abort dependencies are a mechanism for enforcing dependencies between transactions in a workflow [5]. If transaction T' takes a commit dependency on transaction T, then T' will not commit until after T commits. For example, in an insurance claim workflow, a transaction that pays a deposit for an auto repair could take a commit dependency on a previous transaction that accepts the repair shop's estimate.

As a follow-on to Jim's 1981 advice about incorporating transactions into programming languages, in his 2006 talk he mentions transactional memory as a mechanism for simplifying error handling. This is a generalization of the try-catch fault handling model that can potentially help in parallel programming, and which had become unavoidable with the advent of multi-core systems. While research on transactional memory has made progress, challenging problems remain. The same can be said, now, over 15 years later.

In summary, his research advice was to focus on temporal (i.e., multiversion) databases, simplifications of replication as a path to durability, continue looking for a universal workflow abstraction, and find ways to leverage transactions for cleaner and simpler fault handling.

Paper #5: Jim Gray: **A Comparison of the Byzantine Agreement Problem and the Transaction Commit Problem.** Fault-Tolerant Distributed Computing 1986, Lecture Notes in Computer Science 448, Springer 1990: 10-17

The Asilomar Workshop on Fault-Tolerant Distributed Computing in 1986 brought together researchers on distributed computing systems, distributed computing theory, distributed database systems, and fault-tolerant computing. The proceedings remains an excellent reference for many problems in this technical area.

One of the sessions was a discussion of the similarities and differences between the Byzantine Generals Problem which was studied in the distributed computing community and the Atomic Transaction Commit Problem whose solution, two-phase commit, was a core component of distributed databases. Jim wrote this short paper to summarize the result of that discussion. It is still one of the best summaries of that topic.

The problem was originally explained using the following scenario [13, 21]. Some Byzantine army divisions, each led by a general, are planning to attack a city. The generals communicate by sending messengers between them. However, messengers are unreliable (they may be killed) and possibly traitorous (they may alter messages). Some generals also may be traitors. Devise an algorithm whereby all the loyal generals are guaranteed to agree on the same decision, despite the unreliable and traitorous messengers and generals.

It was proved in [21] that you need at least four generals to reach a decision when there is one traitorous general. You might think that three generals are enough, because you can reach a decision with two out of three. However, a traitorous general could send Attack to one general who therefore decides to attack and send Don't Attack to the other general who therefore decides not to attack. Each of the loyal generals incorrectly thinks that a quorum of generals has made the same decision. A loyal general will decide to attack and be unpleasantly surprised that the other two generals do not attack.

This sounds very similar to the atomic commitment problem in transaction processing, where all data managers that participate in a transaction must make the same commit decision. That is, either they all commit or they all abort. However, this correctness criterion is different than the Byzantine Generals problem, where only the non-faulty processes must agree on the decision.

A second difference between the two problems is the type and number of faults that a solution can tolerate. A commit protocol can tolerate many faults, including message loss from a non-failed process. Byzantine protocols work correctly provided less than N/3 of the system's N processes are faulty.

A third difference is that commit protocols never give an incorrect result. Byzantine Agreement protocols (i.e.,

solutions to the Byzantine Generals problem) might give an incorrect result if N/3 or more of the system's N processes are faulty.

A fourth difference is that Byzantine Agreement protocols give an answer within a fixed time bound. Commit protocols don't make this guarantee. In fact, they can't make this guarantee since there are situations where the operational processes don't know whether a failed process committed or aborted and unanimity is required.

Jim's paper lays this all out in detail along with some practical examples. I believe the paper was the first to explain the two problems with the same terminology and highlight their differences precisely.

Paper #6: Hal Berenson, Philip A. Bernstein, Jim Gray, Jim Melton, Elizabeth J. O'Neil, Patrick E. O'Neil: **A Critique of ANSI SQL Isolation Levels.** SIGMOD Conference 1995: 1-10

Two-phase locking prevents the concurrent execution of transactions that have conflicting accesses to the same data. This limits transaction throughput. For this reason, weaker forms of isolation are commonly used, notably, degree 2 isolation, which was described earlier in "Granularity of Locks and Degrees of Consistency in a Shared Data Base." Recall that degree 2 requires that a transaction sets a long duration write lock on data that it updates, and a short duration read lock on data that it reads. Throughput under degree 2 can be as much as three times that of using two-phase locking. That throughput difference translates directly into hardware cost. The cost saving of using weaker isolation levels is a powerful incentive to use it.

To make weaker isolation levels easier to understand, people have given them intuitive names. Degree 2 isolation is commonly called Read Committed, because the locking protocol ensures that a transaction can only read committed data. Degree 3 isolation adds the requirement that a transaction sets a long duration read lock on data that it reads. It is commonly called Repeatable Read, because the locking protocol ensures that a transaction that reads an entity twice will see the same value both times, unless it updated the entity in between.

A SQL database system needs a way for users to specify the isolation level of their SQL statements. Therefore, when defining the ANSI SQL-92 specification [1], the standards committee included keywords for isolation levels and their semantics.

Since not all SQL implementations use locking for concurrency control, the specification could not use concepts like long duration and short duration locks. It needed a specification that admitted non-locking solutions. It settled on defining isolation levels in terms of the following prohibited behaviors:

- Dirty Read – Transaction T1 modifies a row and transaction T2 reads that row before T2 performs a Commit.
- Non-repeatable read – Transaction T1 reads a row, transaction T2 modifies or deletes that row and commits, and then T1 re-reads that row and receives the modified value or discovers the row was deleted.
- Phantom – Transaction T1 reads a set of rows that satisfy a search condition, transaction T2 generates one or more rows that satisfy the search condition, and T1 again reads rows satisfying the same search conditions and obtains a different set of rows.

It then defined four isolation levels:

- Read uncommitted – all the above behaviors are possible.
- Read committed – dirty reads are not possible.
- Repeatable read – dirty reads and non-repeatable reads are not possible.
- Serializable – all three of the above behaviors are not possible.

Although the intent of these definitions is clear, they had some rather surprising implications, which are the main subject of this paper. They illustrate the many pitfalls of trying to explain the behavior of concurrent transactions in intuitive terms. Many of them are rather technical without major practical implications. However, one is quite important. The specification says, "The execution of concurrent SQL-transactions at isolation level SERIALIZABLE is guaranteed to be serializable." Although the specification's intention is clear, that last statement is false.

How could this happen? Consider a database system that uses multiversion concurrency control. When a transaction begins, it is given access to a snapshot of versions that were produced by committed transactions. All the transaction's reads are executed on this snapshot. Therefore, the system avoids dirty reads. Since the snapshot does not change during the transaction's execution, it is unaffected by concurrently executing transactions. Therefore, it ensures repeatable reads and avoids phantoms, at least as they are defined in the specification. Therefore, this hypothetical database system avoids the three prohibited behaviors. Nevertheless, an execution using this mechanism might not be serializable.

For example, consider two transactions T1 and T2 that read the same two rows X and Y from the same database snapshot. Suppose T1 updates row X and T2 updates row Y. Since a transaction's snapshot does not change during its execution, neither transaction reads the updated row produced by the other transaction. Then they both commit. This satisfies the ANSI definition of Serializable. However, in a serial execution of these two transactions, one of them would have read the other one's output, which did not happen in this execution. Therefore, despite satisfying the ANSI definition of Serializable, the execution is not equivalent to a serial execution of the two transactions.

In fact, the database system using the multiversion concurrency control mechanism described above is not just hypothetical. This example execution is allowed by a popular concurrency control protocol called Snapshot Isolation. It works as follows:

- Each transaction is assigned a *commit timestamp* when it commits. It attaches that timestamp to all the versions it wrote.
- When a transaction *T* starts executing it is assigned a *start timestamp*, *st*. When *T* reads an entity, it selects the version with the largest (i.e., latest) commit timestamp less than or equal to *st*.
- When *T* is ready to commit, the system checks the entities that *T* wrote. If any of them were updated by a committed transaction with timestamp greater than *st*, then *T* aborts. This is called *first-writer-wins.* Otherwise, the system assigns *T* a commit timestamp *ct* greater than any it has already assigned to other transactions and attaches *ct* to all versions that *T* wrote.

The first-writer-wins rule ensures that snapshot isolation prevents race conditions. That is, if two transactions read and write the same entity, and both of them read the entity before either of them writes it, then the one that commits first will commit and the other one will abort. In addition, it avoids the three prohibited behaviors of the ANSI definition of Serializable. However, it allows the non-serializable execution of the example above.

There has been much debate about the importance of executions like the one above that satisfy snapshot isolation yet are not serializable. It is not easy to come up with compelling examples of such executions in practice. Hence, some database systems support snapshot isolation and do not offer a truly serializable concurrency control protocol, such as two-phase locking. Apparently, their customers are satisfied with this limitation. In fact, many database system vendors report that most of their users execute their transactions with Read Committed isolation.

This paper led to a series of research papers exploring ways of strengthening snapshot isolation to make it truly serializable, culminating in a paper in 2008 that showed how to do it [4].

Like many research papers, this paper came about due to a happy coincidence. This story was told to me by Hal Berenson, the first author of the paper. Hal was an experienced database system engineer and had recently joined Microsoft to work on its database products. He got very frustrated that multiple teams were implementing serializability incorrectly. To fix the problem, he wrote a document explaining why these teams were wrong and what it meant to be serializable. Pat and Betty O'Neil, who were professors at the University of Massachusetts at Boston and experts in database technology, were spending their sabbatical at Microsoft. Pat happened to wander into Hal's office as he was finishing up the document and asked what Hal was working on. When Hal told him, Pat exclaimed "I think I know why; the language in the standard is wrong." They pulled up the language and Hal was shocked that the description of serializability was indeed incomplete. Jim Melton, the editor of the SQL standard, couldn't find anyone to write the isolation level language for the standard. Finally, he wrote the language himself, trying to avoid using locking terminology. He asked Hal to review it, and Hal didn't catch the problem at the time. Pat suggested that he and Hal collaborate on a paper, and Hal agreed. Hal suggested that since Jim Gray and I were also at Microsoft and were responsible for much of the past research on transactions, he and Pat should see if we wanted to join in. Jim suggested we rope in Jim Melton.

Paper #7: Jim Gray, Pat Helland, Patrick E. O'Neil, Dennis E. Shasha:
**The Dangers of Replication and a Solution.**
SIGMOD Conference 1996: 173-182

This paper explores how best to execute transactions in a database system that runs on a dedicated network and on laptop computers that have only intermittent connectivity. Data is replicated, residing on both host computers and on laptops. This creates scalability challenges since updates must propagate to all replicas.

When this paper was published in the mid-1990's, intermittent connectivity of laptops was common. When used at the office, a laptop computer usually had a reliable high-bandwidth internet connection. However, at home or at a hotel, you connected your laptop to the internet via a dial-up telephone line, with at most 56 Kb/second bandwidth. Everywhere else, a laptop was a standalone device with no network access. It would be at least another decade before fast access was commonly available over Wi-Fi and mobile phone networks.

One approach to the problem of intermittent connectivity was Lotus Notes, released in 1990. Lotus Notes allows updates to a replica of a document on a disconnected computer. When the computer connects to another computer running Notes, the two computers merge their updates to the document. In essence, documents in Notes comprise a replicated database where every replica is independently updatable, and updates to different replicas of a data item are periodically reconciled. The Lotus Notes model of replication came to be known as *update-everywhere* or *multi-master* replication.

The alternative to update-everywhere is the *primary-copy* approach, which allows only one updatable replica of a data item. Updates to the primary are propagated to replicas, either *eagerly* as part of the transaction that updated the primary, or *lazily*, after committing the transaction that updated the primary. With eager propagation, standard transaction synchronization (such as two-phase locking and two-phase commit) ensures that writes to the replicas are applied atomically and in the same order as they were to the primary. With lazy propagation, writes are sent asynchronously to replicas and may arrive in different orders at different replicas, possibly after a long delay. Hence, lazy replication requires additional synchronization to ensure replicas have the same value.

This paper arose from a discussion between Jim and Dennis Shasha, a professor at New York University who was doing some consulting on Wall Street. There, the prevailing opinion was that lazy replication was too weak and using two-phase commit to ensure atomicity of updates to replicas was undesirable because a node failure could make the database unavailable. The question was whether there is an intermediate strategy between these weak and strong synchronization approaches.

To address this question, the paper explores the scalability of the four approaches to updating a replicated distributed database that synchronizes transactions using two-phase locking: update-everywhere with eager or lazy update propagation and primary-copy with eager or lazy update propagation. These techniques were well known when the paper was published. What was new was its analysis of replication's scalability challenges and a proposed approach to primary-copy, lazy replication.

First, consider a system that uses update-everywhere with eager update propagation. In such a system, disconnected computers cannot execute update transactions since they cannot update all replicas. But even when computers are connected to all replicas, there is a serious problem: the system is prone to deadlock when transactions that update the same data execute on different replicas. For example, suppose there are two nodes $N_1$ and $N_2$, both of which have a replica of data item X. Suppose transaction $T_1$ executes at node $N_1$, transaction $T_2$ executes at $N_2$, and both transactions update X. If $T_1$ and $T_2$ execute concurrently, then they are likely to deadlock. To see why, consider the following order of operations:

1. $T_1$ and $T_2$ execute concurrently. That is,
    a. $T_1$ executes on its local copy of X at $N_1$.
    b. $T_2$ executes on its local copy of X at $N_2$.
   At this point, both transactions are still active, waiting for their updates of X to propagate.
2. $T_1$ propagates its write to $N_2$. However, the write is delayed, waiting for $T_2$ to release its lock on X.
3. $T_2$ propagates its write to $N_1$. Here too, the write is delayed, waiting for $T_1$ to release its lock on X. Deadlock!

As the probability of transaction conflict increases and the number of nodes increases, the probability of deadlocks increases by quite a lot. From the above example, it is easy to see why. If the time between the execution of two conflicting transactions is less than the time to propagate a transaction's writes to all replicas, then the transactions are likely to deadlock. The paper presents some formulas that show just how rapidly the deadlock rate will increase as a function of conflict rate and degree of replication.

The second approach is primary-copy replication with eager write propagation. It avoids the deadlock problem because, as explained earlier, writes are applied in the same order at all replicas. Therefore, replication has no effect on the deadlock rate. Thus, it is not surprising that this approach is popular in today's distributed replicated database systems.

The third approach is update-everywhere with lazy write propagation. Here, the problem is that an update transaction may be operating on stale data. For example, reconsider the case of $T_1$ and $T_2$ above. After $T_1$ executes at $N_1$, it may be some time before $N_2$ receives $N_1$'s write. Therefore, if $T_2$ executes after $T_1$, it may be operating on stale data at $N_2$. Since neither transaction reads the result written by the other one, the execution will not be serializable. Writes that propagate lazily have to be reconciled.

To determine if reconciliation is required, each write operation to a replica can include both the old and new value of the data item at the replica where the transaction originally executed. When a write operation is applied to a replica, it first checks that the replica's value is the old value that the transaction observed when it first executed. If so, then it is safe to do the write. If not, then reconciliation is needed, like in Lotus notes. In our example, when applying $T_1$'s write to X at $N_2$, the system checks that the old value of X is the same as it was at $N_1$ before executing $T_1$. If so, then it is safe to apply the write because there was no conflicting update to X at $N_2$ between the time $T_1$ executed at $N_1$ and the time its update is applied at $N_2$. If not, then manual reconciliation is required.

Like the case of update-everywhere with eager write propagation, this case is problematic. Although it avoids the problem of frequent deadlocks, instead it suffers from the problem of frequent reconciliation. So why does this work well in Lotus Notes? The answer is that most updates commute because they are insertions. For example, they send a timestamped message or add a timestamped note to a bulletin board. Neither of these overwrite a previous value, and their timestamps ensure the recipient eventually sees the writes in their intended order. Therefore, reconciliation is not required.

The final case is primary-copy replication with lazy write propagation. Since writes are propagated asynchronously, writes to the same data item might be applied in different orders at different replicas. A common solution is to tag each write with a timestamp or sequence number that indicates the order in which the writes were applied at the primary [25]. A write is applied to a replica only if its timestamp is larger than the maximum timestamp of any previous write that was applied to the replica. This technique, often called *Thomas' Write Rule*, ensures that the replicas of a data item converge to the same value.

In general, lazy write propagation leaves open the possibility that queries on replicas see inconsistent data. For example, suppose transaction $T_3$ updates X, and transaction $T_4$ reads X and updates Y. Consider an execution where $T_4$ reads the value of X written by $T_3$. A node N that stores X and Y might receive $T_4$'s write to Y before it receives $T_3$'s write to X. A query that reads X and Y at N after $T_4$ wrote Y and before $T_3$ wrote X has read an inconsistent database.

The paper offers an analytic model that characterizes the scalability limit of each of the four approaches to replication. The model characterizes deadlock rate and reconciliation rate as a function of the transaction rate, the number of operations per transaction, the execution time of each operation, the number of data items in the database, and the number of nodes. It assumes each node stores a copy of the database and executes transactions at a fixed rate. Thus, each node N that is added to the system has two effects on the workload: its replicas generate work for the other nodes, since they have to propagate their transactions' writes to N; and it adds another stream of transaction executions. This has a quadratic effect on the total workload. It would be more informative to separate the two effects by analyzing the effect of adding replicas without adding more transactions. The resulting formulas would have smaller exponents, making them a little less scary. But they would still be scary enough, and the scalability challenges of the four replication strategies would be unchanged.

In general, a primary-copy lazy propagation scheme is not appropriate for disconnected devices, because they often cannot access the primary copy. As an alternative, the paper concludes by proposing a *two-tier replication* scheme. The system consists of *base nodes*, which contain the entire database and are always connected to each other, and *disconnected nodes*, which usually contain only part of the database and only occasionally connect to a base node. Every data item has a primary copy, which might reside on a base node or a disconnected node. A disconnected node N can execute any transaction T that read and write data stored at N. However, if N does not have the primary copy of some data that T reads, then T runs in *Tentative Mode*.

When N reconnects to the base node, it first discards versions of data written by Tentative transactions, since they will be refreshed after reconciliation with the base node. Then it sends its Tentative transactions and the results of its non-Tentative transactions to the base node. It also accepts replica updates from the base node, all of which are non-Tentative.

After receiving transactions from N, the base node stores the results of N's non-tentative transactions on the base node's local replicas. It then re-runs N's Tentative Mode transactions on primary copies (some of which might reside at N). For each one, if its output differs from the original execution, then it runs an application-specific *acceptance test*. If the acceptance test succeeds, it installs the transaction's updates and returns the result of the acceptance test and transaction to N. If not, it returns a diagnostic message to N.

Disconnected operation is a less common scenario today than it was when this paper was published. However, it remains important. Some sensor systems need to operate disconnected for long periods due to limited available power. Network failures do occur, and when they do, it is important that update transactions can still be executed. Some military applications might need to avoid network communications. Space applications are another possible example. Although the time constants differ in all these cases, the need for mechanisms that address disconnected scenarios remains.

Paper #8: Jim Gray, Leslie Lamport:
**Consensus on transaction commit.**
ACM Trans. Database Syst. 31(1): 133-160 (2006)

In his paper "A Transaction Model" presented earlier in this chapter, Jim described Lampson and Sturgis' two-phase commit protocol, which solves the atomic commitment problem. It ensures the atomicity of a distributed transaction: either all participants in the transaction commit or they all abort. The atomic commitment problem is an example of a consensus problem, a problem in distributed systems where a set of processes must agree on some value that is proposed to them. There are many versions of the problem, depending on the types of failures that the solution must handle, the nature of the consensus, and how the processes communicate. Another consensus algorithm is Paxos [14], invented by another Turing Award winner, Leslie Lamport, and independently as an algorithm called Viewstamped Replication by Brian Oki and Barbara Liskov [18]. On the surface, the two-phase commit and Paxos algorithms seem similar, but they are not identical. To reconcile the differences, Jim and Leslie Lamport collaborated on the above paper. They applied insights from Paxos to two-phase commit, yielding a provably correct and optimized protocol called Paxos Commit.

Earlier, we encountered another consensus problem: the Byzantine Generals Problem. In that problem, the processes need to reach an agreement even if some of the processes experience *Byzantine faults*, where they do not follow the previously agreed upon protocol.

The original two-phase commit protocol by Lampson and Sturgis and most of its commercial implementations do not cope with Byzantine faults. Rather, it assumes that each participant follows the protocol. It also assumes each participant does not lose state information that it wrote to persistent storage. However, the protocol does cope with several types of failure. A participant might fail by stopping (e.g., crashing). Messages might be lost or duplicated. And if a message or stored record is corrupted, then the corruption is detectable.

There are two types of processes in two-phase commit: a *transaction manager* (*TM*) and *resource managers* (*RMs*). The TM drives the execution of the protocol and makes the commit or abort decision. The RMs are the processes that executed the transaction and must all commit or all abort. RMs are usually data managers that read or wrote data on behalf of the transaction, though other types of processes might also be RMs.

In the absence of failure, the Lampson and Sturgis protocol works as follows.

1. When the transaction has finished executing, it asks the TM to commit. The TM sends a PREPARE-REQUEST message to all RMs.
2. To process a PREPARE-REQUEST, each RM does the following:
    a. It writes the transaction's updates to persistent storage.
    b. Then it replies to the TM with a PREPARED message.
   
   Its write to persistent storage ensures that if the protocol fails to finish, it can finish committing the transaction later.
3. If the TM receives a PREPARED reply from all RMs, then it can commit the transaction by writing the commit decision to persistent storage and then sending a COMMIT message to the RMs. If the TM does not receive a PREPARED reply from one or more RMs within its timeout period, then it aborts the transaction. To do this, it writes the decision to persistent storage and sends an ABORT message to all RMs.

In the absence of failures, there are three one-way message delays between the start of the protocol and the time that an RM knows the transaction has committed: the PREPARE-REQUEST from TM to RM, the PREPARED reply from RMs to TM, and the COMMIT notification from TM to RMs.

These message delays adversely affect performance. While the commit protocol plays out, other transactions don't know whether updates by the committing transaction will be committed or undone. Therefore, they need to wait for the commit/abort decision before they can read data that the committing transaction updated. These message delays degrade transaction throughput and latency.

Failures make matters worse. If the TM fails, then some RMs might be blocked, waiting for the TM to recover so it can tell the RMs what was decided (unless they can find this out from other RMs). Transactions that need to read data that the blocked transaction updated are blocked too, since they don't know whether those updates will be committed or undone. If the TM is down for a long time, the impact can be quite serious.

To avoid blocking when the TM fails, protocols have been proposed that use backup TMs in addition to the primary TM. They are usually called "three-phase commit." After receiving all the PREPARED replies and before notifying the RMs that it has decided to commit, the primary TM notifies the backups. If the primary TM fails, the backups can take over. This adds two message delays to the protocol: a COMMIT notification from the primary TM to backup TMs and an acknowledgement from backup TMs to the primary TM.

Since backup TMs might fail and recover, perhaps multiple times, the process of having them take over for the primary TM and reach a commit-or-abort decision is itself a consensus problem. We may not be able to avoid such failures and recoveries. But at least we want to be sure that if the set of operational TMs remains stable for long enough, the operational TMs will reach a unique and durable decision on whether the transaction committed or aborted.

In a sense, three-phase commit incorporates two consensus protocols, one to reach the commit decision and one to ensure the decision is resistant to subsequent failures. The insight in Paxos Commit is that these two consensus rounds can be combined into one by reaching consensus on the prepare decision rather than the commit decision. This ends up saving one message delay. To explain this, we need to backtrack a bit by describing Paxos and how to use it in an atomic commitment protocol.

A key difficulty in developing a consensus algorithm is coping with processes failing and recovering while the algorithm is executing. A good basis for dealing with this difficulty is majority consensus [25]. If a majority of processes reaches a decision, then any future majority of the processes will include one of the processes that reached a decision. A "future majority" might arise because members of the earlier majority failed and other processes that were not part of the earlier majority have recovered and joined the set of operational processes. As long as any future majority honors the decision of the earlier majority, the decision will be stable. For the concept of "future majorities" to work, majorities must be sequenced.

Majority consensus was well known at the time of Gray and Lamport's paper and was used in three-phase commit protocols. However, as Gray and Lamport state, at the time of their paper none of the published protocols provided "a complete algorithm proven to satisfy a clearly stated correctness condition," at least not for the case of asynchronous communication, where a missing message does not imply that the sender has failed. This asynchronous case is problematic because even though the primary TM seems to have failed because it is not responding to messages, it might actually just be slow. If it reappears after another TM has taken over, two TMs now think they are the primary and could reach different decisions.

For the case of synchronous communication, where a missing message implies the sender failed and will not recover, there were provably correct three-phase commit algorithms. They appear in Chapter 5 of Dale Skeen's Ph.D. thesis [19] and in my book with Vassos Hadzilacos and Nathan Goodman [3] (in Section 7.5, which was written by Hadzilacos). However, synchronous communication is an impractical assumption in most distributed systems, so from a practical standpoint, the above quote claiming the lack of provably correct protocols is correct.

There were also provably correct algorithms for consensus, which was known to be a general version of the atomic commitment problem. But descriptions of these algorithms didn't look at their efficiency when used for atomic commitment. The contribution of the Paxos Commit algorithm is that it addressed both correctness and efficiency.

Lamport's Paxos algorithm solves the general consensus problem under the same failure assumptions as two-phase commit. It is general in the sense that the processes, called *acceptors*, can agree on one of many values, not just "commit" or "abort". The process structure includes *acceptors* that cooperate to choose the value and one distinguished acceptor, called the *leader*, which drives the protocol. It is tempting to think of the acceptors as RMs and leader as TM, but this analogy is not exact.

The algorithm proceeds in a sequence of rounds, called *ballots*. Each ballot has a leader, which might differ from ballot to ballot. If a majority of acceptors (which usually includes the leader) accepts the leader's proposed value and the leader confirms that result with a majority of acceptors, then that is the decision. If the leader fails or doesn't receive acceptances of its proposed value from a majority of acceptors, then the ballot is unsuccessful. In that case, one or more other leaders step in to try another ballot with a higher sequence number. This continues until the set of acceptors is stable for long enough for a decision to be reached. After that happens, any future ballots will reach the same decision as the earlier one that was accepted. The algorithm guarantees that if a majority of acceptors remains operational for long enough, the algorithm will reach a decision.

This paper proposes Paxos Commit, an atomic commit protocol that uses Paxos to reach a commit or abort decision. The process structure is similar to (but not identical to) two-phase commit. It uses a Paxos leader to initiate the protocol. The leader sends a PREPARE-REQUEST message to RMs, each of which runs an instance of Paxos to fault-tolerantly report its decision to prepare. Each RM sends its PREPARED message to the acceptors, each of which forwards the PREPARED message to the leader. For each RM, when the leader receives a PREPARED from a majority of acceptors, it knows the RM has prepared. When the leader knows that all RMs have prepared, the transaction is committed.

Each acceptor is participating in a Paxos instance for all RMs. Therefore, instead of forwarding each RM's PREPARED message to the leader separately, an acceptor can wait until it receives PREPARED from all RMs and then send just one PREPARED to the leader.

In Paxos Commit the majority of the acceptors that accept PREPARED messages from RMs substitutes for the TM writing the decision to storage. The acceptors function like backup TMs in two-phase commit.

The paper is interesting for many reasons. It explains the difference between Paxos and two-phase commit. It proposes the Paxos Commit protocol, which shows how to use Paxos to implement two-phase commit. It shows that two-phase commit is the trivial version of Paxos Commit that tolerates no faults, in the sense that even a single fault can cause it to block (though it won't cause two-phase commit to give an incorrect result). It explains several optimizations of Paxos Commit, which are applicable to other atomic commitment protocols too. And perhaps most importantly, it introduces a different process structure than previous atomic commitment protocols, in that it treats each RM's transition to the prepared state as a consensus problem.

The different process structure of Paxos Commit addresses the main inefficiency in an atomic commitment protocol, namely, the number of message delays to commit a transaction. Paxos Commit has four message delays: Request-Prepared from leader to RMs, Prepared from RM to acceptors, Prepared from acceptors to leader, and Committed from leader to RMs. This is one message-delay fewer than three-phase commit, because RMs send Prepared directly to acceptors, which function as backups, rather than first sending Prepared to the primary TM, which makes the commit decision and then sends Commit to backups and waits for a majority of backups to acknowledge their receipt.

(The paper says there are five message delays, because it includes the first message to the leader to initiate the protocol. We omit this message here since it is common to Paxos Commit and two-phase commit.)

The paper also suggests an optimization where each acceptor sends its Prepared messages to all RMs, instead of sending them to the leader. Each RM can independently make the commit decision after it receives a majority of Prepared messages from every RM. Although this adds more messages, it avoids the message delay of requiring the leader to send the Commit decision to the RMs.

To eliminate another message delay, the paper suggests having each RM spontaneously send its Prepared message when it has persisted the transaction's updates, rather than waiting to receive a Prepared-Request from the leader. This suggestion is problematic unless RMs know that the transaction has terminated. Usually, they know this because they received a Prepared-Request. If an RM prepares spontaneously without knowing the transaction terminated, then it might receive another update from the transaction after it has prepared. Since the RM has announced it has prepared, the transaction might commit before that last update is processed, which would be an error. To avoid this problem, most systems wait for the transaction to terminate before initiating the atomic commit protocol, which begins by sending a Prepared-Request to all RMs.

The paper explains the above ideas in more detail than the above summary. It includes an analysis of the number of messages exchanged in each variation of Paxos Commit. Its appendix has a formal specification of the protocol in Lamport's TLA+ specification language, "for the most committed readers" (as the paper says). It's a worthwhile read.

Lamport invented the Paxos Commit algorithm. However, he and Jim collaborated on its relationship to the two-phase commit protocol, and hence on this paper. Lamport summarized the history of the paper on his publications page, here:

https://lamport.azurewebsites.net/pubs/pubs.html#paxos-commit

There are many later publications that describe protocols that draw on the ideas in Paxos Commit. Most of them integrate the protocol with a storage service. See [7, 10, 11, 17 , 26, 28].

**ACKNOWLEDGMENTS**

I am very grateful to Pat Helland for having asked me to write this article and for having recommended the eight papers to cover. I thank Leslie Lamport and Dennis Shasha for their substantial help with the sections on *Consensus on Transaction Commit* and *The Dangers of Replication and a Solution*, respectively.